\documentclass[preprint,preprint,12pt]{elsarticle}

\usepackage{amssymb}
\usepackage{xfrac}
\usepackage{amsmath}

\journal{Thin-Walled Structures}

\begin{document}

\begin{frontmatter}



\title{The “reversed von Kármán/Brazier effect” and a compendiary analytical solution} 


\author[1]{Ida Mascolo\corref{cor1}}%

\author[2]{Igor Orynyak}
\author[1]{Federico Guarracino}

\cortext[cor1]{ida.mascolo@unina.it}
\affiliation[1]{organization={Department of Structural Engineering, University of Naples "Federico II"},
            addressline={Napoli}, 
            country={Italy}}
\affiliation[2]{organization={Department of Applied Mathematics at National Technical University, Kiev Polytechnic Institute},
            addressline={Kiev}, 
            country={Ukraine }}

\begin{abstract}
The goal of the present study is to offer some accurate analytical formulae for the evaluation of the here defined “reversed von Kármán/Brazier’s effect”, which may appear at first sight anomalous and has previously been noticed and successively explained using both Finite Element analyses and approximate analytical formulations. The proposed analytical formulae, based on the previous contribution of one of the present authors and on the Donnell’s equations for cylindrical shells can be helpful in emphasizing the possible impact that some arrangements may have in various fields and especially in offshore engineering and can assist design calculations.
\end{abstract}

\begin{highlights}
\item The \emph{reversed von Kármán/Brazier's effect} for cylindrical shells under bending is introduced and defined.
\item The effects of an imposed or restricted ovalisation of the cross section are analytically evaluated.
\item The validation of the proposed formulae is performed with respect to experimental findings and numerical analyses.
\end{highlights}


\end{frontmatter}



\section{Introduction}
\label{sec1}
\begin{figure}
    \centering
    \includegraphics[width=0.5\linewidth]{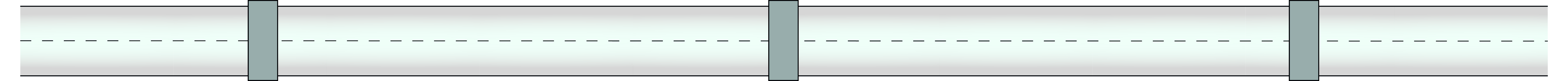}
    \caption{Buckle arrestors.}
    \label{Fig_1}
\end{figure}

The ovalisation of the circular cylindrical shells under bending is a well-known phenomenon  first described and analysed by von Kármán~\cite{vonKármán1911} and successively by Brazier~\cite{brazier1927flexure}, that naturally occurs in many problems and has a certain relevance in offshore engineering. In this framework it is known that in order to prevent a propagating collapse failure in deepwater pipelines, it is practical to add buckling arrestors, such as thick-wall rings, at regular intervals along the pipeline (see Figure~\ref{Fig_1}). In the event of an accident, a number of these stiffeners will reduce the amount of pipe damage~\cite{palmer1975buckle}. As pipe support collars, these stiffeners are also utilized with pipes that are installed by J-Lay procedures. It has been noticed in the past that the use of stiffeners and buckling arrestors can inhibit the well-known von Kármán--Brazier effect, which is the cause of the ovalisation, giving origin to an unexpected change in the strain and stress distribution in the pipe walls~\cite{guarracino2008analysis,guarracino2009effects}.
Moving from these experimental findings, the goal of the current paper, following a number of previous contributions provided by the last of the present authors over the years~\cite{guarracino2003analysis, guarracino2008analysis, guarracino2009effects, guarracino2011simple,guarracino2007considerations}, is to present a simple but comprehensive mathematical formulation of this problem on the basis on the Donnell’s equations for cylindrical shells. The proposed treatment summarises a phenomenon that, to the best of authors’ knowledge, has never been pointed out and can be useful in several fields, from mechanical to civil engineering and biomechanics. More specifically, the presented formulae allow to evaluate the effects of stiffeners and buckle arrestors on the bending of pipelines, which can happen especially at the installation stage~\cite{guarracino1999refined}, without the necessity to turn to intricate finite element analyses.

\section{Experimental findings} \label{Sec_2}

\begin{figure}
    \centering
    \includegraphics[width=1\linewidth]{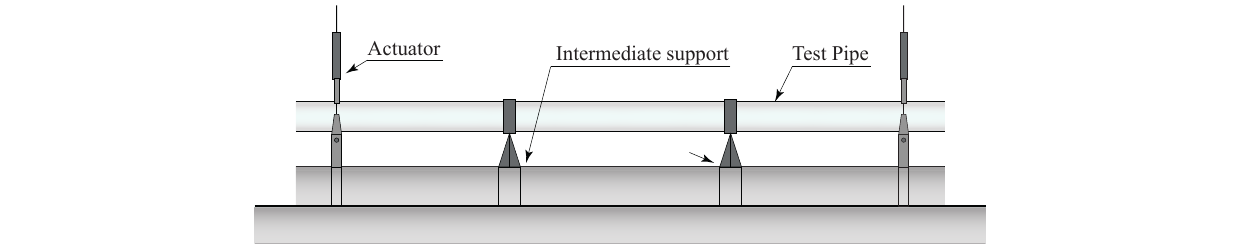}
    \caption{Typical four-point bending test arrangement.}
    \label{Fig_2}
\end{figure}

In the present Section a brief resumè of experimental findings already discussed in previous works~\cite{guarracino2008analysis, guarracino2009effects,guarracino2011simple} is given.
In the industrial practice, a segment of a circular cylindrical shell is typically tested under only bending loads on the assumption that the test specimen will deform in accordance with the basic bending beam theory.  Primarily, this implies that the application of a pure bending moment will result in maximal tensile and compressive strains that are equal in size while the material is still elastic.  Figure~\ref{Fig_2} depicts a typical test setup for a medium-sized pipe: the core part of the tested pipe is expected to be subjected only to bending action, with no, or at most very little, shear or axial stresses, as the test rig applies a four-point bending condition.
It has been conventional to assume that during the loading process the axial strains have identical values in tension and compression and that the strains can be calculated directly from the curvature or the vertical displacements of the central section of the pipe because the pipe is thought to be a simple structural element and the Eulero-Bernouilli beam theory is thought to hold true.  In experiments where the pipe was loaded to the point of local buckling, the ultimate strain values were typically estimated from measurements of the deflections. Notably, strain gauges have been systematically fastened to the test pipe in order to measure axial strains directly only in the past twenty-five years. 
However, some experiments on $152mm$ diameter pipe conducted back in $1985$ to ascertain the maximum curvature to which the pipe may be deformed~\cite{cit8} also employed strain gauges to measure axial and circumferential strains. The pipe was reinforced at the supports and at the loading locations by very thick steel collars. In order to prevent localized loading on the pipe wall, the collars were machined to fit extremely snugly around the pipe. As a result, the pipe was completely prevented from ovalizing at each of these sites. In the test rig's design, it was expected that a center test section located at around $5D$ from the collars would be sufficient to guarantee that end effects from the loading circumstances would fade away. The axial strains were found to be rather consistent along the test section's length, but the averaged values of the compressive and tensile strains were found to deviate significantly from one another. The apparent discrepancy between the measured strains and the expected values in relation to the simple bending theory was not investigated at the time, even after it was confirmed that the strain gauges were correctly positioned and the instrumentation was operating as intended. 

An inquiry into the root cause of the anomaly was successively conducted due to the significance of the test results in determining the permitted amounts of strain for lined pipes. This is covered in detail in~\cite{guarracino2008analysis,guarracino2009effects}  and special emphasis was devoted to the constraint arrangements in testing. Experimental evidence was found that compressive stresses were generally greater than the tensile one under bending if ovalisation was prevented, as is the case with applied stiffeners of buckling arrestors, whereas the tensile stresses were generally greater than the compressive stress if some form of localized ovalisation was imposed. The phenomenon was confirmed with the help of a number of finite element models designed to accurately represent the conditions in bending tests or in pipelines that undergo cross-sectional changes while being bent\cite{guarracino2003analysis,guarracino2009effects}.

\section{Analytical treatment} \label{Sec_3}
\subsection{The case of imposed ovalisation} \label{Sec_2.1}

As said before, the experimental evidence was that, differently from simple flexure theory, maximum tensile stresses were greater than the compressive stresses if some form of localized ovalisation was imposed, while maximum compressive stresses were found greater than the tensile ones under bending if ovalisation is prevented.

\begin{figure}
    \centering
    \includegraphics[width=1\linewidth]{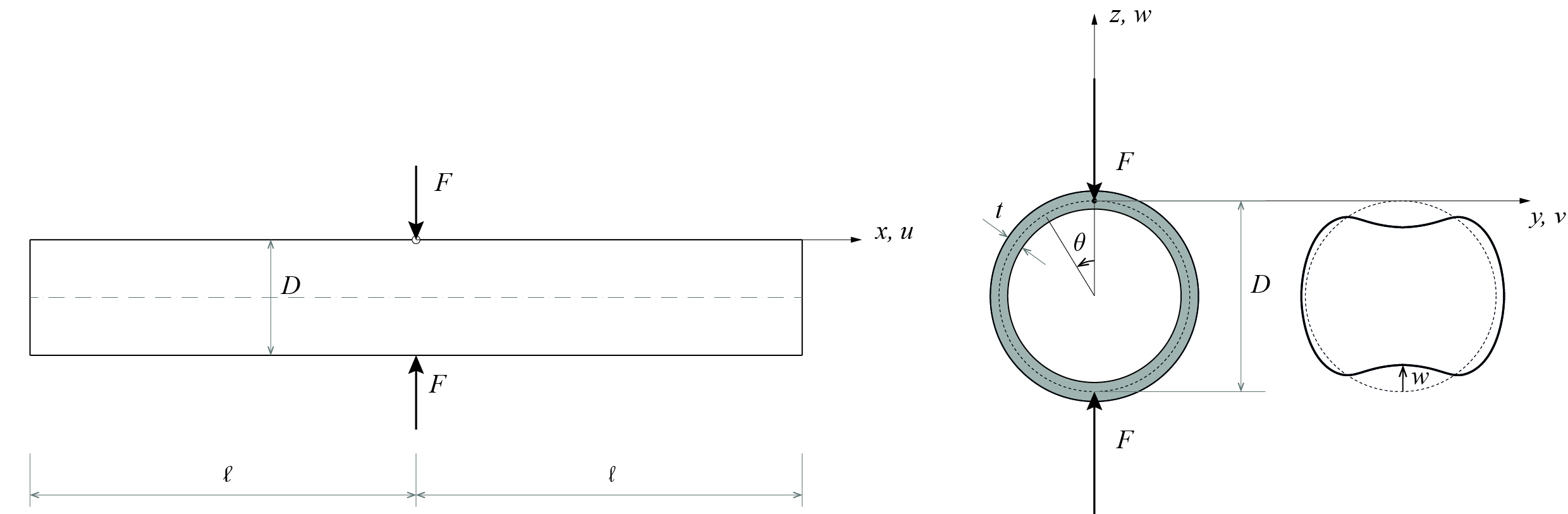}
    \caption{Circular cylindrical shell loaded by two opposite inward forces at the half-length section: coordinate system and notations.}
    \label{Fig_3}
\end{figure}

The first case can be analysed by imagining the superposition to the symmetric bending of a  stress status deriving from the application of two opposite forces directed towards the center at a section of the pipe, see Figure~\ref{Fig_3}.  
In fact, it has been shown, by means of Finite Element analyses and approximate analytical methods~\cite{guarracino2008analysis, guarracino2011simple} that the stress status deriving from such a loading condition, superposed to that of simple bending, can give an accurate account of the experimental findings. 

This problem is here addressed by referring to the Donnel’s equations and to the beam analogy as presented by Calladine~\cite{calladine1977thin,calladine1983theory}. In fact, with reference to the coordinate system and displacements shown in Figure~\ref{Fig_3}, the problem of a circular cylindrical shell loaded by two opposite inward forces can be decomposed into the superposition of several distinct problems, by expressing the load per unit length of circumference,  $p$, in terms of a Fourier series as follows 
\begin{equation}
    \label{eq1}
p = {p_o} + {p_n}\sum\limits_{n = 2}^\infty  {\cos \left( {n\theta } \right)}  = \frac{F}{{\pi R}} + \frac{{2F}}{{\pi R}}\sum\limits_{n = 2}^\infty  {\cos \left( {n\theta } \right)} ,\;\quad n = 2,4, \ldots 
\end{equation}

\noindent being $n$ an even natural number.

The first addendum of Equation~\eqref{eq1} represents a uniformly distributed radial loading, namely a hydrostatic pressure applied along the circumference at the mid-plane. This type of loading causes a radial displacement that can be expressed~\cite{calladine1977thin} as
\begin{equation}
    \label{eq2}
{w^I}(x) = \frac{{F{\mu ^3}}}{{4\pi D\delta }}{e^{ - \frac{x}{\mu }}}\sin \left( {\frac{x}{\mu } - \frac{\pi }{2}} \right)
\end{equation}
\noindent and at the point of force application it results
\begin{equation}
    \label{eq3}
w_0^I = \frac{{F{\mu ^3}}}{{4\pi D\delta }}\,, 
\end{equation}
\noindent being
\begin{equation}
    \label{eq4}
\mu  = \sqrt[4]{{\frac{{{D^2}{t^2}}}{{12\left( {1 - {\nu ^2}} \right)}}}}\quad \quad {\rm{and}}\quad \quad \delta  = \frac{{Et{\mu ^4}}}{{{D^2}}}\,, 
\end{equation}
where $E$ and is the Young’s modulus of the material. The second addendum of Equation~\eqref{eq1} represents the sum of   terms of a Fourier-series with the same amplitude . The asymmetric loading conditions in which the cylindrical shell is loaded by a single n-component of the Fourier-series~\eqref{eq1}, can be analyzed by representing the cylindrical shell by the superposition of two parallel-coupled structural surfaces, one that experiences bending strains and the other that experiences stretching strains~\cite{calladine1977thin,calladine1983theory}. These two surfaces carry the applied loading unevenly and the contest between the purely-inextensional (short wave) and the stretching-dominated (long-wave) behaviour is controlled by a dimensionless half-wavelength parameter, i.e.
\begin{equation}
    \label{eq5}
   \lambda  = \frac{{\sqrt t \ell }}{{\sqrt {{{\left( {D/2} \right)}^3}} }} \,. 
\end{equation}

As a result, both in short-wave (i.e., $\lambda<0.5$)  and in long-wave (i.e., $\lambda>1$) condition, the cylindrical shell can be effectively simplified using a one-dimensional beam model based on the Winkler foundation analogy, where the stiffness,  $m$, of the foundation represents the resistance of the shell to the longitudinal deformation. 
\begin{figure}
    \centering
    \includegraphics[width=1\linewidth]{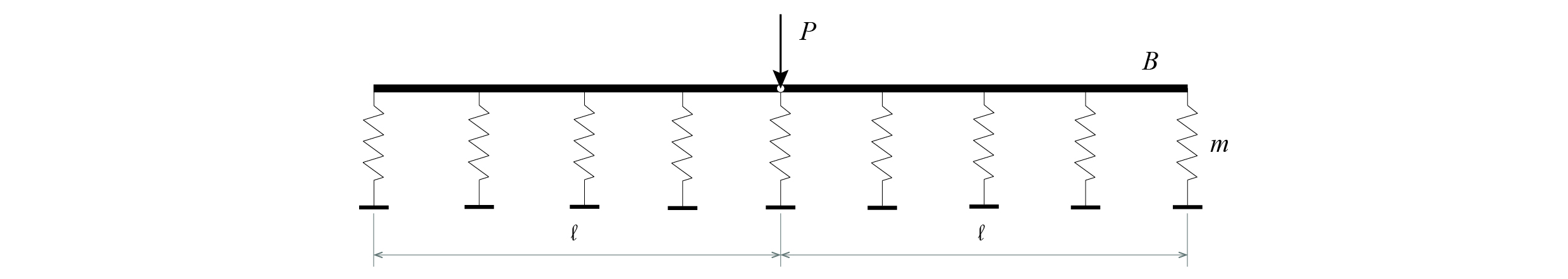}
    \caption{Beam-on-elastic foundation analogy.}
    \label{Fig_4}
\end{figure}

In the intermediate cases (i.e.,  $0.5<\lambda<1$), the above analogy does not hold true in principle and a proper combination of short- and long-wave solutions needs to be investigated~\cite{calladine1977thin,calladine1983theory}. A general procedure for combining such solutions has been proposed by one of the present authors~\cite{orynyak2018efficient,oryniak2021application}.

In the long-wave-dominated cases, which characterize several engineering problems (a typical $D/t$ ratio in offshore engineering ranges between $16$ and $32$) and are the primary object of the present study, the cylindrical shell can be analysed by the equivalent beam resting on an elastic foundation shown in Figure~\ref{Fig_4}. The flexural rigidity of the beam is
\begin{equation}
    \label{eq6}
   B = \frac{{\pi Et{D^3}}}{{8\left( {1 - {\nu ^2}} \right){n^4}}}\,, 
\end{equation}
\noindent while the stiffness of the foundation is
\begin{equation}
    \label{eq7}
m = \frac{{8\pi \delta {{\left( {{n^2} - 1} \right)}^2}}}{{{D^3}}}\,, 
\end{equation}
\noindent and the central point load intensity is
\begin{equation}
    \label{eq8}
P = 2F\,. 
\end{equation}
The vertical displacement of the elastic supported beam, $w$, is~\cite{hetenyi1946beams}
\begin{equation}
    \label{eq9} 
    w\left( x \right) = \frac{{P\,\Psi }}{{2m}}{e^{ - \Psi \,x}}\left( {\cos  \frac{\Psi\,x}{b} + \sin \frac{\Psi\,x}{b}} \right)\,, 
\end{equation}
\noindent where
\begin{equation}
    \label{eq10} 
    \psi  = \sqrt[4]{{\frac{m}{{4B}}}}\,, 
\end{equation}
and $b$ is half-wavelength in the longitudinal direction that can be assumed equal to half of the total length of the cylindrical shell, $\ell$. By summation of the series to infinite, the deflection under the force is 
\begin{align}
    \label{eq11} 
   w_0^{II} &= w(0) = \frac{{2\;{3^{3/4}}P\left( {1 - {\nu ^2}} \right)}}{{\pi E{\rm{ }}R}}{\left( {\frac{R}{t}} \right)^{5/2}}\sum\limits_{n = 2,4}^\infty  {\frac{n}{{{{\left( { - 1 + {n^2}} \right)}^{3/2}}}}} \nonumber \\&\approx \frac{1}{{1.244}}\frac{{P\left( {1 - {\nu ^2}} \right)}}{{E{\rm{ }}R}}{\left( {\frac{R}{t}} \right)^{5/2}}\,. 
\end{align}
Therefore, the resulting total radial displacement at 
$\theta  = \pi /2$ is
\begin{equation}
    \label{eq12} 
w_0^{} = w_0^I + w_0^{II} = \frac{{P\left( {1 - {\nu ^2}} \right)}}{{{\rm{E }}R}}{\left( {\frac{R}{t}} \right)^{5/2}}\left( {\frac{1}{{1.244}} + \frac{1}{{4.774}}\frac{t}{{R{{\left( {1 - {\nu ^2}} \right)}^{3/4}}}}} \right)\,. 
\end{equation}
The displacement component in the axial and in the circumferential direction, $u$ and $v$ , respectively, can be easily derived by noticing that the long-wave approximation is characterized by the following strain conditions
\begin{equation}
    \label{eq13} 
{{\varepsilon _\theta } = \frac{2}{D}\frac{{\partial v}}{{\partial \theta }} + \frac{{2w}}{D} = 0\, ,}\qquad{{\gamma _{x\theta }} = \frac{{\partial v}}{{\partial x}}} + \frac{2}{D}\frac{{\partial u}}{{\partial \theta }} = 0\,, 
\end{equation}
\noindent so that
\begin{equation}
    \label{eq14} 
u =  - \frac{D}{{2{n^2}}}\frac{{\partial {w_n}}}{{\partial x}}\cos n\theta \,, \quad v =  - \frac{{{w_n}}}{n}\sin n\theta \,. 
\end{equation}

\noindent Finally, the strain in the longitudinal direction is 
\begin{equation}
    \label{eq15} 
{\varepsilon _x} = \frac{{\partial u}}{{\partial x}} =  - \frac{D}{{2{n^2}}}\frac{{{\partial ^2}{w_n}}}{{\partial {x^2}}}\cos{n\theta } \,. 
\end{equation}

By virtue of Equation~\eqref{eq15}, the deformation and the stress status to be added to the bending solution in case of imposed ovalisation is readily obtained.  

\subsection{The reversed von Kármán/Brazier's effect}\label{subsec 3.1}
The second case, which is the most relevant in the present study, is the one in which the ovalisation is prevented. This case appears less intuitive but can be easily explained by making reference to what is here defined as a sort of a \emph{reversed von Kármán/Brazier effect}. In fact, it is known that the cross section of thin-walled beams deforms during bending and that this deformation affects the bending strength, giving occurrence to a flattening instability phenomenon. This phenomenon was firstly described and modelled by von Kármán in the case of curved tubes and, successively, by Brazier and Chwalla in the case of originally straight tubes~\cite{brazier1927flexure}.

\begin{figure}
    \centering
    \includegraphics[width=1\linewidth]{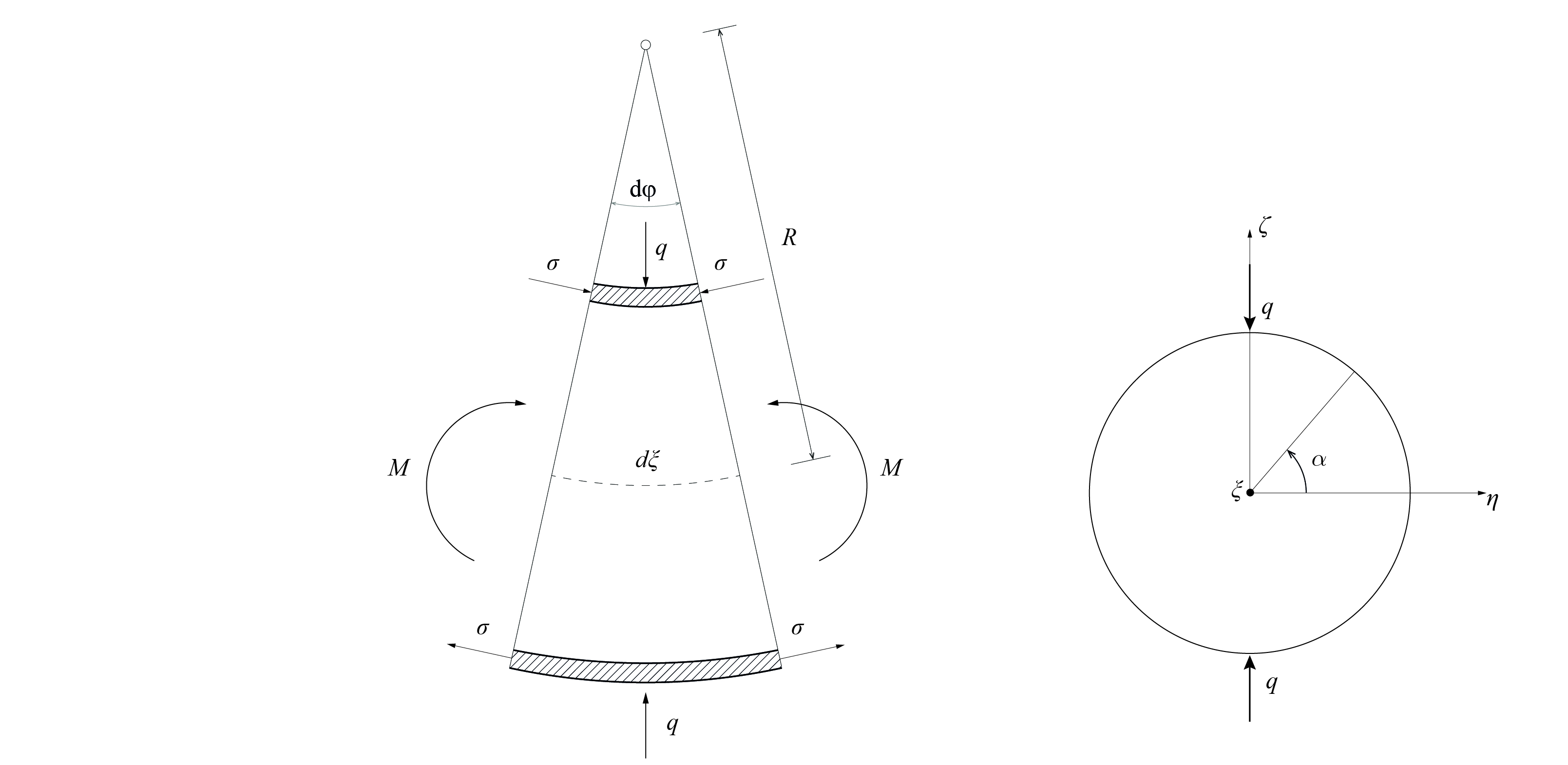}
    \caption{Axial stresses and resulting ovalising pressure.}
    \label{Fig_5}
\end{figure}

In~\cite{guarracino2008analysis} a very straightforward modelling of the von Kármán effect was presented by taking into account the non-linear behaviour of the cross-section of the originally straight tube subject to the ovalising pressure that derives from bending. The results substantially coincided with those obtained by Brazier and it was also shown that a further simplification could be attained by assuming that the originally round section becomes elliptical in shape. The ovalising pressure q acting on the unit of area of the tube wall in the direction normal to the neutral axis (see Figure~\ref{Fig_5}) can be expressed as
\begin{equation}
    \label{eq16} 
q = {\chi ^2}Et\varsigma  = {\chi ^2}Et\frac{D}{2}\sin \alpha  \,, 
\end{equation}

\noindent where $\sigma$ is the stress normal to the cross section and $\chi$ is the curvature of the longitudinal axis of the ring section
\begin{equation}
    \label{eq17} 
\chi  = \frac{1}{R} = \frac{M}{{EI}} = \frac{{2\sigma }}{{ED}} \,. 
\end{equation}

\noindent The nomenclature of the remaining geometrical parameters can be derived from Figure~\ref{Fig_5}. 

\noindent The ovalisation induced by the pressure q is a function of the curvature of the longitudinal axis of the cylindrical shell and can been measured by the shortening of the vertical diameter of the cross section, that is 
\begin{equation}
    \label{eq18} 
\Delta {D^q} = D - 2w_0^{} \,, 
\end{equation}
\noindent where
\begin{equation}
    \label{eq19} 
w_0 = \frac{{{\chi ^2}{D^5}\left( {1 - {\nu ^2}} \right)}}{{32{t^2}}} \,. 
\end{equation}
It can be thus imagined that any prevented ovalisation, that is the here defined \emph{reversed von Kármán/Brazier effect}, corresponds to a counter pressure deriving from Equation~\eqref{eq16} applied to the pipe wall that can be thought equivalent to two opposite forces as shown in Figure~\ref{Fig_3} but reversed in sign. The difference between maximum and compressive stresses will be obviously opposite with respect to the case of imposed ovalisation.

By equating the formulae~\eqref{eq12} and~\eqref{eq19}, it is obtained that the corresponding force $P$ is

\begin{equation}
    \label{eq20}
    {P_{eq}} = \frac{{{\rm{E}}{\mkern 1mu} {R^4}{\chi ^2}}}{{\left( {\frac{1}{{1.244}} + \frac{1}{{4.774}}\frac{t}{{R{{\left( {1 - {\nu ^2}} \right)}^{3/4}}}}} \right)}}{\left( {\frac{t}{R}} \right)^{1/2}}\,.
    \end{equation}

\section{Study cases}\label{Sec 4}

On the bases of the formulation presented in the previous Sections, a steel cylindrical shell with the following properties,  Young's modulus $E=208000MPa$, Poisson's ratio $\nu=0.3$, yielding stress $\sigma_0=450MPa$ , diameter $D=608.6mm$, wall thickness $t=18.9mm$, is taken into consideration. This pipe was originally the object of some experimental tests reported in~\cite{guarracino2008analysis,guarracino2009effects}.

First, the experimental results from testing arrangement reported in~\cite{guarracino2008analysis, guarracino2009effects,guarracino2011simple}, which is essentially the one of Figure~\ref{Fig_2}, with a distance between the actuators and the intermediate supports of $2000mm$ and a distance between the intermediate supports of $2750mm$ is examined. 

\begin{figure}
    \centering
    \includegraphics[width=1\linewidth]{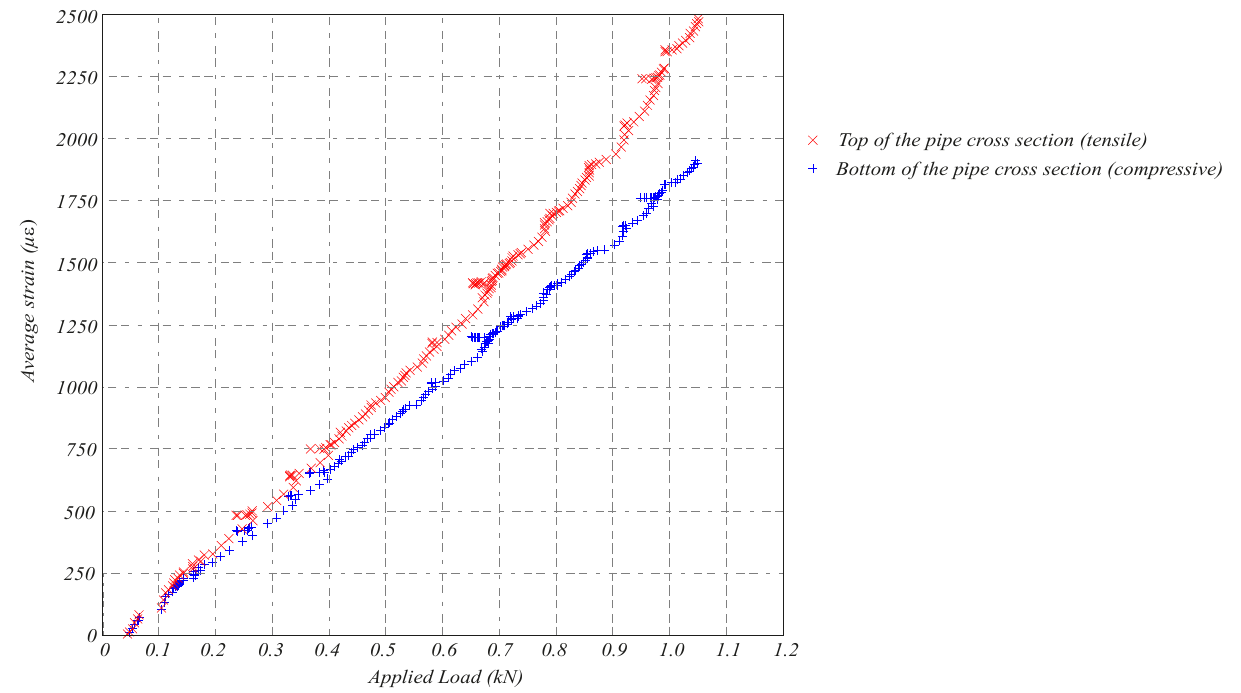}
    \caption{Averaged strain along the top and bottom of the pipe bend test specimen (bottom strain reported with reversed sign).}
    \label{Fig_7N}
\end{figure}

The readings from the longitudinal strain gauges applied at mid-span are plotted in Figure~\ref{Fig_7N} against the applied loading. 
For a value of the applied load of $1MN$, the longitudinal strain at the top and at the bottom of the pipe yielded by Navier's formula under bending is
\begin{equation}
    \label{eq21}
    \varepsilon_x^{bend} = \left|\frac{M}{EI} \frac{D}{2}\right|= 1742.31 \mu\varepsilon\,,
    \end{equation}
while formula~\eqref{eq15} gives, under the approximation that the applied force and the reaction of the supports are supposed to be opposite to each other. 
\begin{figure}
    \centering
    \includegraphics[width=1\linewidth]{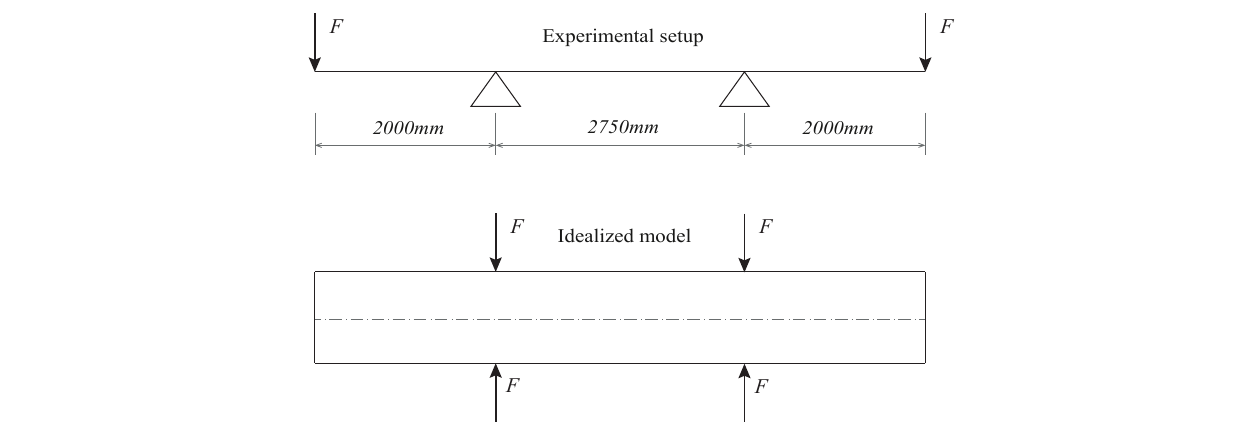}
    \caption{Experimental setup: the case of imposed ovalisation}
    \label{Fig_8}
\end{figure}
At the mid-span, that is at $\sfrac{2750}{2}\, mm$ from the intermediate supports, the longitudinal strain deriving from the imposed ovalisation results $\varepsilon_x^{oval}=305.25 \mu\varepsilon$, so that at the top of the pipe it is
\begin{equation}
    \label{eq22}
    \varepsilon_{x,top}^{tot}=\varepsilon_x^{bend}+\varepsilon_x^{oval}=2047.56 \mu\varepsilon
    \end{equation}
\noindent and at the bottom
\begin{equation}
    \label{eq22}
    \varepsilon_{x,bot}^{tot}=-\varepsilon_x^{bend}+\varepsilon_x^{oval}=-1437.06 \mu\varepsilon\,,
    \end{equation}
\noindent with a theoretical ratio between top and bottom longitudinal strains equal to $1.42$ versus an experimental one of $1.28$. The difference can be obviously ascribed to the fact that, differently from what is implied by the formulae in Subsection~\ref{subsec 3.1}, the forces in the experimental setup are $2m$ apart from one another. Nevertheless, the framework is able to capture the order of magnitude of the phenomenon deriving from the imposed ovalisation. 

The case of restrained ovalisation, that is the \emph{reversed von Kármán/Brazier effect}, is now analysed by assuming that the same pipe is subject to a bending moment, $M$, apt to attain a longitudinal stress at the top and at the bottom of the cross section of $440MPa$, slightly below the yield value. According to Equation~\eqref{eq20} the equivalent force capable of preventing the ovalisation is $27478.5N$ which, by virtue of Equation~\eqref{eq12} implies a shortening of the vertical diameter of $0.662mm$. With acceptable approximation, it is here assumed that the second moment of inertia of the cross section of the pipe does not depend on the curvature $\chi$ of the longitudinal axis. This is due to the fact that for the problem at hand, it can be considered admissible to couple the linear treatment presented in Section~\ref{subsec 3.1} with the non-linear expressions~\eqref{eq16} and~\eqref{eq19}, as it will be proven by the following numerical assessment.
\begin{figure}
    \centering
    \includegraphics[width=1\linewidth]{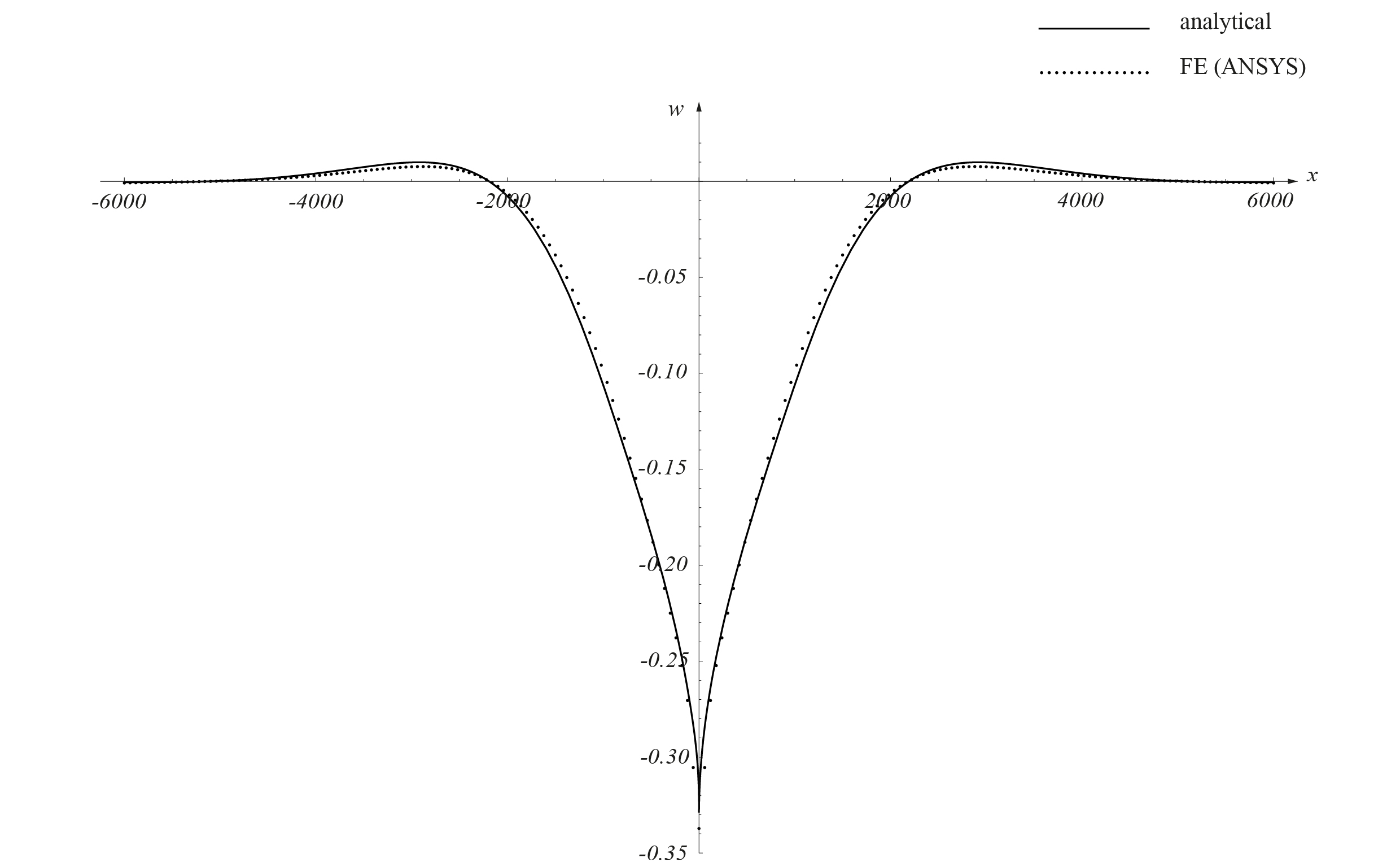}
    \includegraphics[width=1\linewidth]{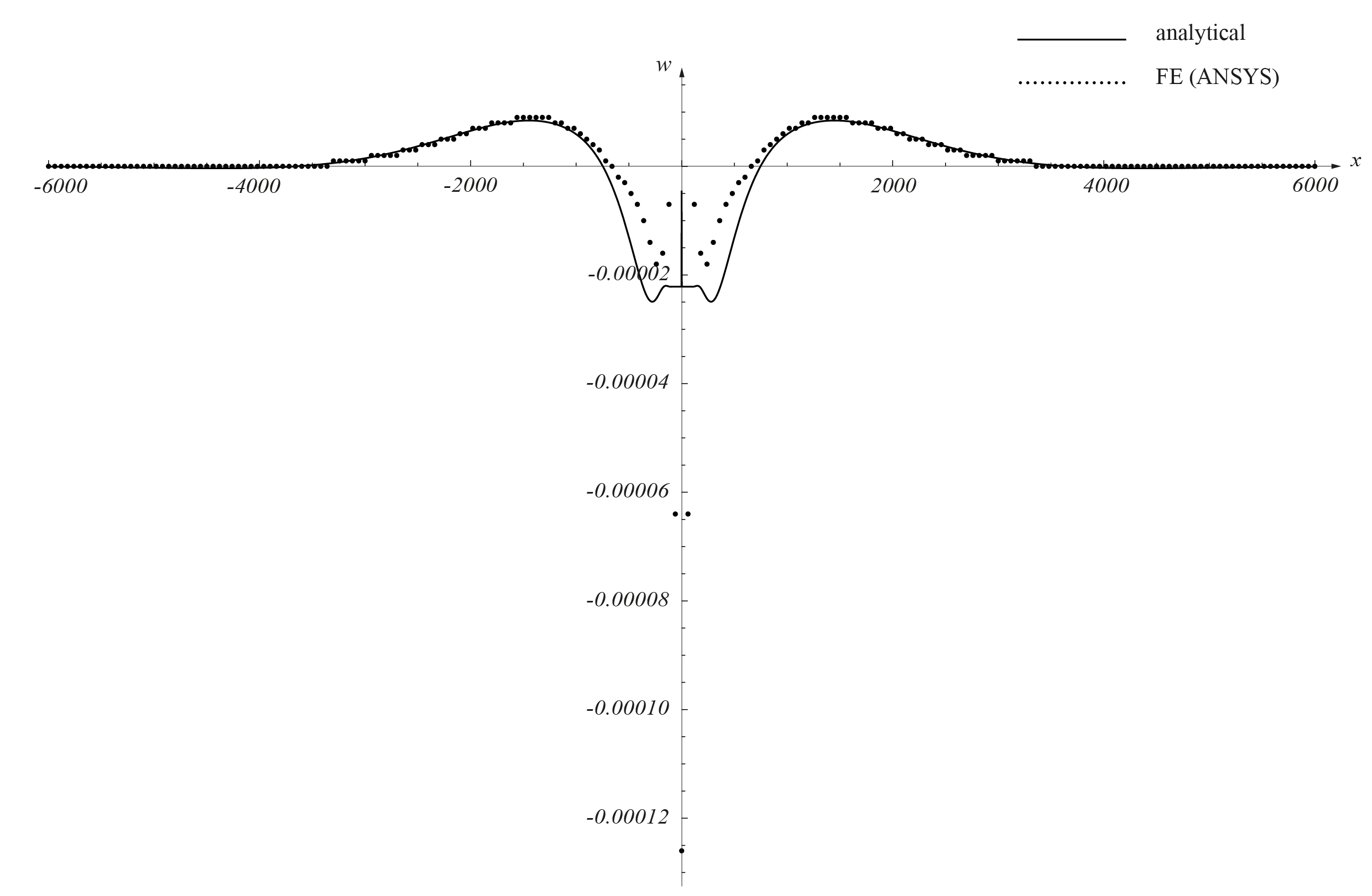}
    \caption{Vertical displacement (top) and Longitudinal strain (bottom) of the points at at the top of the cross section along the axis of the cylindrical shell for a force of $27478.5N$ applied at $x=0$.}
    \label{Fig_6}
\end{figure}

Figure~\ref{Fig_6} shows the vertical displacement and the longitudinal strain of the points at the top of the cross section for the opposite equivalent forces of $27478.5N$ that results in very quite agreement with the results from the FE analysis performed by means of the commercial package Ansys~\cite{Ansys}.

\begin{figure}
    \centering
    \includegraphics[width=1\linewidth]{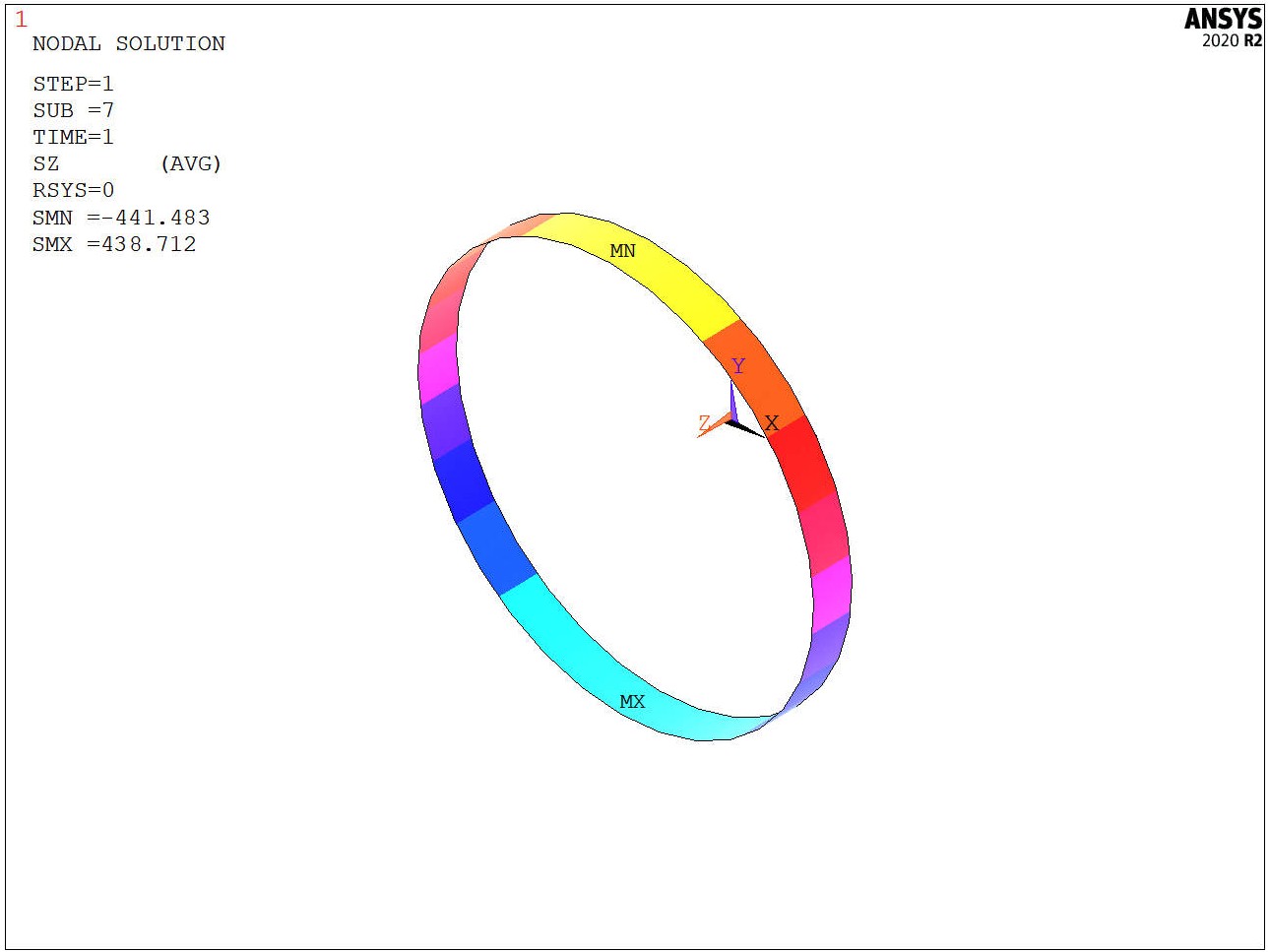}
    \caption{Longitudinal stress at the section located $1.8m$ from the section with ovalisation prevented.}
    \label{Fig_11}
\end{figure}

At a distance of $1.8m$ from the constrained section it is found, in fact, that
\begin{equation}
    \label{eq27}
    \varepsilon_{x,top}^{tot}=2107.84 \mu\varepsilon \,, \quad
    \varepsilon_{x,bot}^{tot}=-2122.93 \mu\varepsilon
    \end{equation}
\noindent and
\begin{equation}
    \label{eq28}
    \sigma_{x,top}=439.40 \mu\varepsilon \,, \quad
    \sigma_{x,bot}=-440.60 \mu\varepsilon\,,
    \end{equation}
\noindent that are in very good agreement with the FE analysis (see Figure~\ref{Fig_11}). 
\section{Conclusions}
The \emph{reversed von Kármán/Brazier’s effect}, i.e. the deviation of strain and stress status of cylindrical shells from pure bending when the ovalisation of the cross section is either prevented or imposed has been defined and analysed making reference to analytical formulae based on the Donnell’s equations. The proposed analytical formulae have been assessed with reference to previous experimental findings and numerical analyses and, apart from their theoretical relevance, can be helpful in emphasizing the possible impact that some arrangements may have in various fields and especially in offshore engineering, by assisting design calculations.



\bibliographystyle{elsarticle-num} 
\bibliography{02_Bib}



\end{document}